\documentclass[useAMS,usenatbib]{mn2e}
\pdfoutput=1
\usepackage{graphicx}
\usepackage{latexsym}

\title{Effects of X-ray irradiation and disk flaring on the [Ne~{\sc ii}]~12.8$\mu$m emission from young stellar objects.}

\author[E. Schisano, B. Ercolano, M. G\"udel]{E. Schisano$^{1,2}$, B. Ercolano$^{3,4}$, M. G\"udel$^{5}$\\
$^{1}$Universit\`{a} degli Studi  di Napoli "Federico II",Corso Umberto I, I-80138 (NA), Italy \\
$^{2}$INAF - Osservatorio Astronomico di Capodimonte, Via Moiariello 16, I-80131 Napoli, Italy\\
$^{3}$Institute of Astronomy, Madingley Road, Cambridge, CB3 0HA, UK\\
$^{4}$Department of Physics and Astronomy, University College London, WC1E 6BT, UK\\
$^{5}$Institute of Astronomy, ETH Zurich, 8093 Zurich, Switzerland}

\begin{document}

\pagerange{\pageref{firstpage}--\pageref{lastpage}} \pubyear{2009}

\maketitle

\label{firstpage}

\def\mnras{MNRAS}
\def\apj{ApJ}
\def\aap{A\&A}
\def\apjl{ApJL}
\def\apjs{ApJS}
\def\araa{ARA\&A}

\begin{abstract}
The [Ne~{\sc ii}] fine-structure emission line at 12.8$\mu$m has been detected in several young stellar objects (YSO) spectra. This line is thought to be produced by X-ray irradiation of the warm protoplanetary disk atmospheres, however the observational correlation between [Ne~{\sc ii}] luminosities and measured X-ray luminosities shows a large scatter. Such spread limits the utility of this line as a probe of the gaseous phase of disks, as several authors have suggested pollution by outflows as a probable cause of the observed scatter. 
In this work we explore the possibility that the large variations in the observed [Ne~{\sc ii}] luminosity may be caused instead by different star-disk parameters. 
In particular we study the effects that the hardness of the irradiating source and the structure (flaring) of the disk have on the luminosity and spectral profile of the [Ne~{\sc ii}]~12.8 $\mu$m line. We find that varying these parameter can indeed cause up to an order of magnitude variation in the emission luminosities which may explain the scatter observed, although our models predict somewhat smaller luminosities than those recently reported by other authors who observed the line with the {\it Spitzer} Space Telescope. Our models also show that the hardness of the spectrum has only a limited (undetectable) effect on the line profiles, while changes in the flaring power of the disk significantly affect the size of the [Ne~{\sc ii}] emission region and, as a consequence, its line profile. In particular we suggest that broad line profiles centred on the stellar radial velocity may be indicative of flat disks seen at large inclination angles.

\end{abstract}

\begin{keywords}

\end{keywords}

\section{Introduction} 
\indent Young solar mass stars are surrounded by $\sim10^{-2}$M$_{\odot}$ of dust and gas distributed in a circumstellar disk which is the birthplace of planetary systems. 
In recent years efforts have been made to study the inner regions ($\la 10$~AU) of these disks in order to constrain gaseous planetary system theories, and to understand the physical processes that characterise the warm and chemically active regions of protoplanetary disk atmospheres.
Accurate theoretical modelling of the spectral energy distribution (SED) of young stars (see reviews by Dullemond et al. 2007 and Natta et al. 2007), coupled with observations in the near and mid-infrared carried out with the $Spitzer$ Space Telescope have greatly improved our understanding of the dust component in the inner disks. On the other hand, direct studies of the gaseous component of these disks are more challenging, even though the gas exceeds the dust by over two orders of magnitude in mass. So far molecular line emission, like the CO fundamental and overtones \citep{Najita2007}, the rovibrational transition of water \citep{Carr2008,Salyk2008},  and H$_{2}$ \citep{Herczeg2002}, have been used as diagnostics of regions with temperatures around 1000-2000~K. 

The [Ne~{\sc ii}]~12.8$\mu$m line (from now on [Ne~{\sc ii}]) has been suggested as a new probe of gas in the potentially planet forming regions of the inner disks (e.g. Glassgold et
al. 2007, Hollenbach \& Gorti, 2009). The ionised region of the disk where this line is formed is heated mainly by X-rays from the central star (Glassgold et al. 2007, Meijerink et al. 2008, Ercolano et al. 2008a, Glassgold, Ercolano \& Drake, 2009; but see also Gorti \& Hollenbach, 2008). 
Indeed, observations from space, with the {\it Spitzer} Infrared Spectrometer (IRS) \citep{Pascucci2007,Lahuis2007,Espaillat2007}, and from the ground, with MICHELLE and TEXES at Gemini North (Herczeg et al., 2007; Najita et al., 2009) and VISIR at VLT (Van Boekel et al. 2009), have successfully detected the [Ne~{\sc ii}] line in tens of young systems. 
The mean observed line luminosity is comparable to the predicted value from disk
models, but the scatter of the observed values around this mean spans over three
orders of magnitude (e.g. G\"udel et al. 2009). If the [Ne~{\sc ii}] line is to be used as a useful diagnostic of the gaseous phase of the inner disks its origins must be well understood. In particular if the line is formed in the disk
by irradiation of the gas by X-rays from the central star then one naturally
expects to find a correlation between line luminosity and X-ray luminosity for
the observed sources. Such correlation is indeed observed (Pascucci et al 2007,
G\"udel et al. 2009), however, for objects with the same X-ray luminosity the [Ne~{\sc ii}] line luminosities span over at least one order of magnitude even after removing those sources with known outflows, for which it can be argued that a large contribution to the line comes from the outflow itself (e.g. T Tau, van Boekel et al. 2009). Understanding the origin of this scatter is extremely important if useful information is to be extracted from future observations of this line. 
 For example, one should expect that disk structural properties, like presence of holes or different degree of flaring, not considered in earlier models have an effect on the  [Ne~{\sc ii}]  emitting region. Moreover, one relevant question is whether the [Ne~{\sc ii}] line originates in the bound layers of the disk irradiated mainly by X-rays or whether it includes a contribution from a photoevaporative outflow (e.g. Alexander, 2008). 
In this paper we will attempt to answer the latter question by exploring the possibility that the scatter observed in [Ne~{\sc ii}] line luminosities for a given X-ray luminosity may be due to details of the irradiating field or to the disk structure (i.e. degree of flaring). We will also produce theoretical line profiles and will compare them to available high resolution observations (e.g. Herczeg et al. 2007, Najita et al. 2009). 

We use the models of Ercolano et al. (2009) to calculate [Ne~{\sc ii}] line luminosities and line profiles for irradiating spectra of varying hardness. We also show the results from additional models that were run to investigate the effects of disk flaring on the line luminosities and profiles. The paper is organised as follows. The modeling strategy is outlined in Section 2; Section 3 contains a review of our results, while our final conclusions and discussion are given in Section~4.

\section{Modelling strategy and methods}

We investigate the effects of two parameters on the [Ne~{\sc ii}] emission region of protoplanetary disks: (i) hardness of the irradiating spectrum; (ii) flaring of the disk. 

\begin{figure}
\resizebox{\hsize}{!}{\includegraphics[width=0.48\textwidth, trim = 25mm 125mm 15mm 25mm, clip]{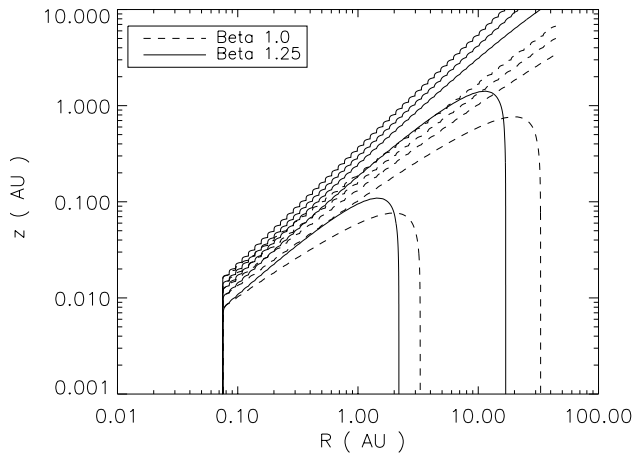}}
\caption{Isodensity curves of the two input disk models used in this work to explore the impact that different degree of flaring has on [Ne~{\sc ii}] line luminosity. The black solid lines show the 'flat' model ($\beta$ = 1.0) and dashed lines show the 'flared' model ($\beta$ = 1.25). 
The plotted contours refer to increasing densities from 10$^{5}$, at higher heights, to 10$^{13}$ g cm$^{-3}$ deeper into the disk with steps of 10$^{2}$ cm$^{-3}$.}
\label{Fig:IsodensityModels}
\end{figure}

\subsection{Hardness of the irradiating spectrum} We calculated fine structure [Ne~{\sc ii}] line luminosities using the temperature and ionisation structure obtained by Ercolano et al. (2009)
with the MOCASSIN code for EUV+X-ray irradiated protoplanetary disks. We refer to Ercolano et al. (2008a, 2009) for details of the disk models, and to Ercolano et al. (2003, 2005, 2008b) for details of the code used. The atomic database included opacity data from Verner et al. (1993) and Verner \& Yakovlev (1995), energy levels, collision strengths and transition probabilities from Version 5.2 of the CHIANTI database (Landi et al. 2006, and references therein) and the improved hydrogen and helium free-bound continuous emission data of Ercolano \& Storey (2006). In particular the Ne$^+$ atomic data include collisional strengths of Saraph \& Tully (1994) and the transition probabilities of Blackford \& Hibbert (1994). We briefly
summarise here the model input parameters: we used a stellar mass of 0.7~M$_{\odot}$, radius 2.5 R$_{\odot}$ and effective temperature of 4000 K and an initial disk mass of 0.027~M$_{\odot}$
with an outer radius of $\sim$500AU\footnote{Note that only the inner 50~AU of the disk are considered here}. The MOCASSIN models are coupled to a hydrostatic equilibrium
routine which ensures that the effects of irradiation on the disk structure are taken into account self-consistently. 
 Line luminosities computed without such routine result to be underestimated at worst by a factor of $\sim$ 2.   
The models were irradiated with synthetic spectra extending from the EUV to
the X-ray spectral region (13.6 eV-10 keV) where the spectra were calculated using the multi-temperature plasma prescription described by Ercolano et al. (2009) and were computed using the PINTofALE IDL software
suite \citep{Kashyap2000}, adopting the solar chemical composition (Grevesse \& Sauval, 1998), the atomic data from the CHIANTI compilation (Landi et al. 2006) and ion populations from
\citet{Mazzotta1998}. The transmittance of the synthetic spectra through neutral screens of varying thickness (as expected to be found in the circumstellar environment of pre-main sequence stars,
e.g. G\"udel et al., 2007, 2008) was calculated assuming neutral hydrogen column densities in the range $N_{H} = 10^{18}-10^{22}$ cm$^{-2}$, with increasing columns resulting
in increasingly harder spectra, as the soft radiation is absorbed out. We present [Ne~{\sc ii}] luminosities and line profiles for the Ercolano et al. (2009) models labeled in their work as FS0H2Lx1,
FS18H2Lx1, FS19H2Lx1, FS20H2Lx1, FS21H2Lx1 and FS22H2Lx1, which consist of models irradiated by spectra screened by columns of 0., $10^{18}$, $10^{19}$, $10^{20}$, $10^{21}$, and
$10^{22}$cm$^{-2}$, and an X-ray luminosity after the screen of L$_{X}$(0.1-10~keV) = 2$\times$10$^{30}erg/s$ (``base luminosity'' hereafter; please refer to Table~1 in Ercolano et al. (2009) for a full description of the model parameters). 
Keeping L$_{X}$(0.1-10 keV) constant {\it after the screen} by normalization of the {\it absorbed spectrum} was done here in order to isolate the effects of the incoming radiation field hardness alone.
Our prescription results in the unscreened models to have a total ionising luminosity
 L$_{\rm tot}$(13.6 eV-10 keV) of approximately twice the X-ray luminosity and the moderately screened models (N$_H$~=~10$^{18}$  cm$^{-2}$) to have a total ionising luminosity 1.2 times the base luminosity.
 We ran four additional models with screens of 0 and 10$^{21}$cm$^{-2}$ and X-ray luminosities an order of magnitude higher and lower than the base luminosity in order to ensure that any scatter produced by varying screen
thickness at fixed L$_{X}$ = 2$\times$10$^{30}erg/s$ would also be reproduced for higher or lower X-ray luminosities.    

\subsection{Disk Flaring} In order to investigate the effect of disk flaring on [Ne~{\sc ii}] line luminosities, we used a simple analytical description of the disk structure, as given by Robitaille et al (2006). 
The density distribution is described by a fixed power-law decline of the surface density in the radial direction coupled with a gaussian relationship along the z-direction:

\begin{eqnarray} 
\rho(R,z) = \rho_{0}\left(1 - \sqrt{{R_{\star} \over R}}\right)\left({R_{\star} \over R}\right)^{\alpha}\exp{\left(-{{z^2} \over {2h^2(R)}}\right)}
\end{eqnarray} 

\noindent where $\rho_{0}$ is a scale factor fixed by the disk total mass and $h(R)$ is the disk pressure scaleheight parametrized by a power-law: 
:
\begin{eqnarray} 
h(R) = h_{in}\left({R \over R_{in}}\right)^{\beta}
\end{eqnarray} 

\noindent where R$_{in}$ and h$_{in}$ are, respectively, the dust destruction radius and the hydrostatic equilibrium scaleheight at this radius.
Different values of $\beta$ produce disk models with different degree of flaring, with higher $\beta$ producing higher flaring. We adopted $\beta$ = 1.0 and  $\beta$ = 1.25 to build models for a 'flat' and 'flared' disk, respectively,
 and we calculated the corresponding  [Ne~{\sc ii}] luminosities and line profiles.  Such values are also 
used as limits for the $\beta$ parameter in the set of SEDs of \cite{Robitaille2006}, typically used in 
the literature to fit real observations. 
For either density structure we  also chose R$_{in}$ corresponding to a dust sublimation temperature of 
1600 K, i.e. R$_{in}\sim$0.08 AU with the previous central star parameters, z$_{factor}$ = 1 from model 
comparison with the disk structure of \cite{Ercolano2009} and $\alpha$ = $\beta$ + 1 \citep{Whitney2003a,
Whitney2003b} to preserve the surface density distribution power law scaling $\Sigma {\rm (R)} \sim {\rm R}^{-1.5}$ 
of the minimum solar nebula model of Hayashi (1981).
The other stellar and disk parameters were chosen to be the same as those used by the Ercolano et al. (2009) models described above. 

Figure \ref{Fig:IsodensityModels} shows the isodensity curves of the two disk models with different flaring powers. The flat model has a scaleheight reduced typically by a factor of about 3-4 between cylindrical radii from 1 to 10 AU when compared to the flared model. In order to isolate the effects of disk flaring only we keep the density structures fixed and do not impose hydrostatic equilibrium on these models. 

\subsection{Line profiles}

We computed line profiles from all our models assuming Keplerian rotation at several disk inclinations. The profiles were computed by summing up the contributions from each cell which produces a Doppler broadened profile given by 

\begin{equation} 
\Phi(R,\theta,z,v) = {{I(R,z)} \over {\sqrt{2\pi}v_{th}(R,z)}} \exp{\left(-{{(v - v_{los}(R,\theta))^2} \over {2v_{th}(R,z)^2}}\right)}
\end{equation} 
where $I(R,z)$ is the power emitted by each cell and $v_{th}$ is the local thermal velocity of the Ne$^+$ ions equal to  
\begin{equation} 
v_{th}(R,z) = \sqrt{{3k_{B}T(R,z)} \over {m_{{\rm Ne}}m_{{\rm H}}}}
\end{equation} 
and $v_{los}(R,\theta)$ is the cell velocity projected along the line of sight.
Given the temperature structure of the disk, the width of the thermal broadening in each cell is $\le$6 km\,s$^{-1}$.
We assumed that the midplane is optically opaque to the [Ne~{\sc ii}] line radiation so only contributions from one side of the disk are taken into account when computing the profile. 
In fact, adopting the dust cross section from  Weingartner and Draine 2001 of R$_V$ = 5.5 extinction law, generally used to reproduce the interstellar extinction in molecular clouds, 
$\sigma_{12.8\mu m}$  = 1.6 $\times 10^{-23}$ cm$^2$ / H  and $\sigma_{15.5\mu m} = 1.4 \times 10^{-23}$ cm$^2$ / H, we find that  optical depths of the order of ~1 are reached for hydrogen column density of about 6-7 x 10$^{22}$ cm$^{-2}$. 
Such column density are reached, for all inclination $\le$ 60, at heights of $\le$ 0.1 AU at a radial distance of 1 AU and $\sim$1 AU at a distance of 10 AU, deeper in the disk than the Ne emitting region shown in Figure \ref{Fig:Emiss}. In the case of almost edge-on systems the column density along the line of sight could be large enough to attenuate the observed line intensities, this is not taken into account in this work, but should be in future comparisons with real observations. Finally, we ignore scattered light contribution to the profile, which is negligible in the Weingartner \& Draine (2001) dust model that has an albedo of order 0.01 at 12.8 $\mu$m, but could be significant for different dust models. 
The resolution of the line profiles we computed  is $\lambda/\Delta \lambda \sim150000$, since we summed all together the emission from cells whose $v_{los}$ falls in bins wide 2~km\,s$^{-1}$. Moreover, we have convolved the line profiles with a gaussian function with FWHM of 10~km\,s$^{-1}$ to degrade them  to a resolution $\lambda/\Delta \lambda \sim30000$, comparable with the high-resolution spectroscopic observations.

\begin{figure}
\includegraphics[width=0.5\textwidth, trim = 25mm 125mm 15mm 25mm, clip]{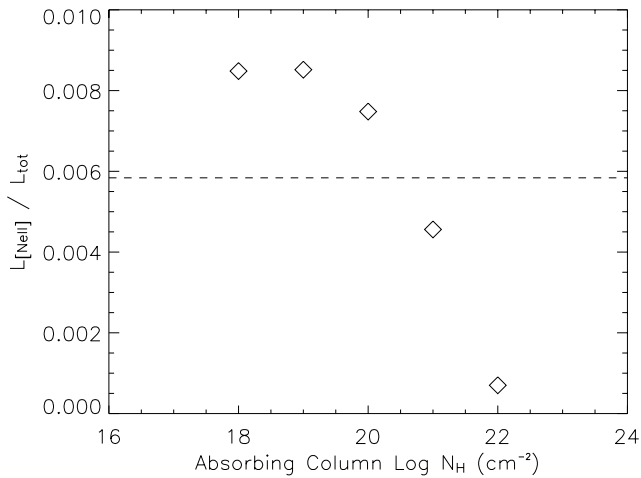}
\caption{L([Ne~{\sc ii}])/L$_{\rm tot}$ as a function of the screening column of the irradiating spectrum. The dashed line indicates the value for the unscreened spectrum. }
\label{Fig:NH}
\end{figure}

\begin{figure}
\includegraphics[width=0.5\textwidth, trim = 25mm 125mm 15mm 25mm, clip]{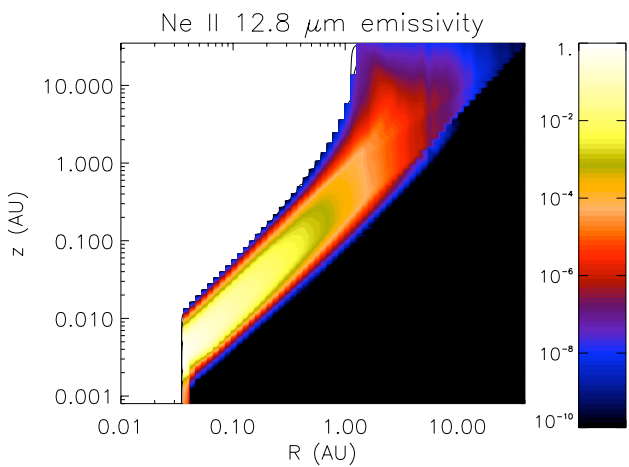}
\caption{Two-dimensional distribution of the [Ne~{\sc ii}] emissivity in arbitrary units for a hydrostatic equilibrium disk model irradiated by an unscreened spectrum.
}
\label{Fig:Emiss}
\end{figure}

\section{Results}

In this section we present the line luminosities and profiles obtained from the models described above. Monte Carlo errors on the given line luminosities were obtained by comparing multiple runs of the same model and are about 5\% for the lines in question.    As in Sect.2.3 we will consider the [Ne~{\sc ii}] line luminosity response to
(i) changes in the shape of the irradiating spectrum and (ii) changes in the disk density distribution (flaring).  

\subsection{Hardness of the irradiating spectrum} 

Table \ref{Table1a} shows the values of [Ne~{\sc ii}]~12.8 $\mu$m  and [Ne~{\sc iii}]~15.5 $\mu$m line luminosities for models irradiated by the multi-temperature thermal spectrum of Ercolano et al. (2009) screened by columns of circumstellar material of thickness 0., 10$^{18}$, 10$^{19}$, 10$^{20}$, 10$^{21}$, 10$^{22}$cm$^{-2}$ (see their Figure~2). 

Figure~\ref{Fig:NH} shows L([Ne~{\sc ii}])/L$_{\rm tot}$ as a function of the screening column. The dashed line indicates the value predicted in the case of unabsorbed source. 
From this figure and from the values in Table~\ref{Table1a} it appears that EUV radiation is less efficient at producing [Ne~{\sc ii}] than the (soft) X-rays. 
In fact, even if the unscreened models have a total ionising luminosity (X-Rays+EUV) which is double than the X-ray only luminosity, they only produce a moderate increase in L([Ne~{\sc ii}]).   This can be easily understood when one considers that [Ne~{\sc ii}] is most efficiently produced in the low ionization layer (just above the molecular layer, see Glassgold et al., 2007; Glassgold, Ercolano \& Drake, 2009), where a fast charge exchange with the abundant neutral H atoms quickly transforms most doubly ionised Ne into singly ionised. The charge exchange coefficient from singly ionised to neutral Ne is instead extremely slow, conspiring to produce high abundance of Ne$^+$ in this warm layer. Instead, EUV radiation ($> 21.6$eV and $< 100$eV) can only produce Ne$^+$ by standard ionisation of neutral Ne and is only efficient in the lower density upper atmosphere where an H~{\sc ii} region like layer is formed. However, some of the Ne$^+$ produced in this region will be ionised to Ne$^{2+}$ and the lower neutral H fractions in this region will result in a much less efficient charge exchange process.   A spatial map of the [Ne~{\sc ii}] emission for the unscreened model is shown in Figure \ref{Fig:Emiss}.
 
Furthermore, the models shown in Table~\ref{Table1a} also suggest that the hard X-ray region is inefficient at the production of [Ne~{\sc ii}] compared with the soft X-ray region. Indeed
models irradiated by hard spectra with N$_H$ ~$\ge$~10$^{21}$ cm$^{2}$  show a steep decline in the [Ne~{\sc ii}] luminosity, even after forcing the X-ray luminosity (0.1-10~keV) to stay constant.
This effect is dominated by the fact that models illuminated by a harder X-ray spectrum result to be colder (see discussion in Ercolano et al. 2009), the hard X-ray penetrate a region that is denser and cooler and therefore the resulting [Ne~{\sc ii}] flux is reduced. Cooler models also result in flatter disks which subtend a smaller solid angle to the source of radiation. 

The hardness of the spectrum has a noticeable effect on the [Ne~{\sc iii}]~15.5 $\mu$m/[Ne~{\sc ii}]~12.8 $\mu$m ratio. This ratio is higher for softer spectra
due to the fact that doubly ionised Ne in these models can be produced in the upper layers of the disk where the neutral hydrogen fraction is lower and therefore charge exchange less efficient.
Observations of the  [Ne~{\sc iii}] 15.5$\mu$m (from now on [Ne~{\sc iii}]) line are still rather sparse although so far low values of the [Ne~{\sc iii}]/[Ne~{\sc ii}] ratio have been inferred (e.g. 0.06, Lahuis et al., 2007), indicating that according to our models, at least for the few objects in question, a soft (EUV-rich) irradiating spectrum is not expected. 

Our results show that changes in the irradiating spectrum can give an order of magnitude scatter (an extra factor of $\sim$2 is introduced when one considers that an unscreened source may contribute comparable ionising luminosity in the EUV band that would not be accounted for in the typical L([Ne~{\sc ii}]) vs L$_X$ relation). 

Figure \ref{Fig:IntegEmiss} (left panel) shows the surface-integrated emission  for some of the models discussed in this Section. We find that in all cases half of the total
[Ne~{\sc ii}] emission comes from the inner regions of the disk, $\la$ 3-4 AU, so that the different spectra do not significantly affect the emission region for this line.
Table~\ref{Table2} lists the half-luminosity and 90\% luminosity radii for models of various screens.  
As the extent of the emission region does not significantly vary with spectral hardness, the line profiles obtained are also roughly invariant. As an example, in Figure~\ref{Fig:lineFS0H0} we show the [Ne~{\sc ii}] profile for the unscreened model seen almost face on (10$^{\rm o}$ inclination) and almost edge on (80$^{\rm o}$ inclination), which have Full Width Half Maximum (FWHM) of $\sim$6-8 and $\sim$25-31~km\,s$^{-1}$  respectively.  

\subsection{Disk Flaring}
Table \ref{Table1b} also shows the results for the L([Ne~{\sc ii}]) for models with flared ($\beta$ = 1.25) and flat ($\beta$ = 1.0) disks. A casual inspection of this table reveals
roughly an order of magnitude increase in the relative luminosity of the [Ne~{\sc ii}] line when the flaring is changed from $\beta$ = 1.0 to $\beta$ = 1.25. This is easily understood when
considering that a more flared disk would subtend a larger solid angle to the irradiation source. 

\begin{figure*}
\begin{center}
\centering
\begin{minipage}{.98\linewidth}
\hspace{-5mm}
\includegraphics[width=90mm,origin=c, trim = 25mm 125mm 15mm 25mm, clip]{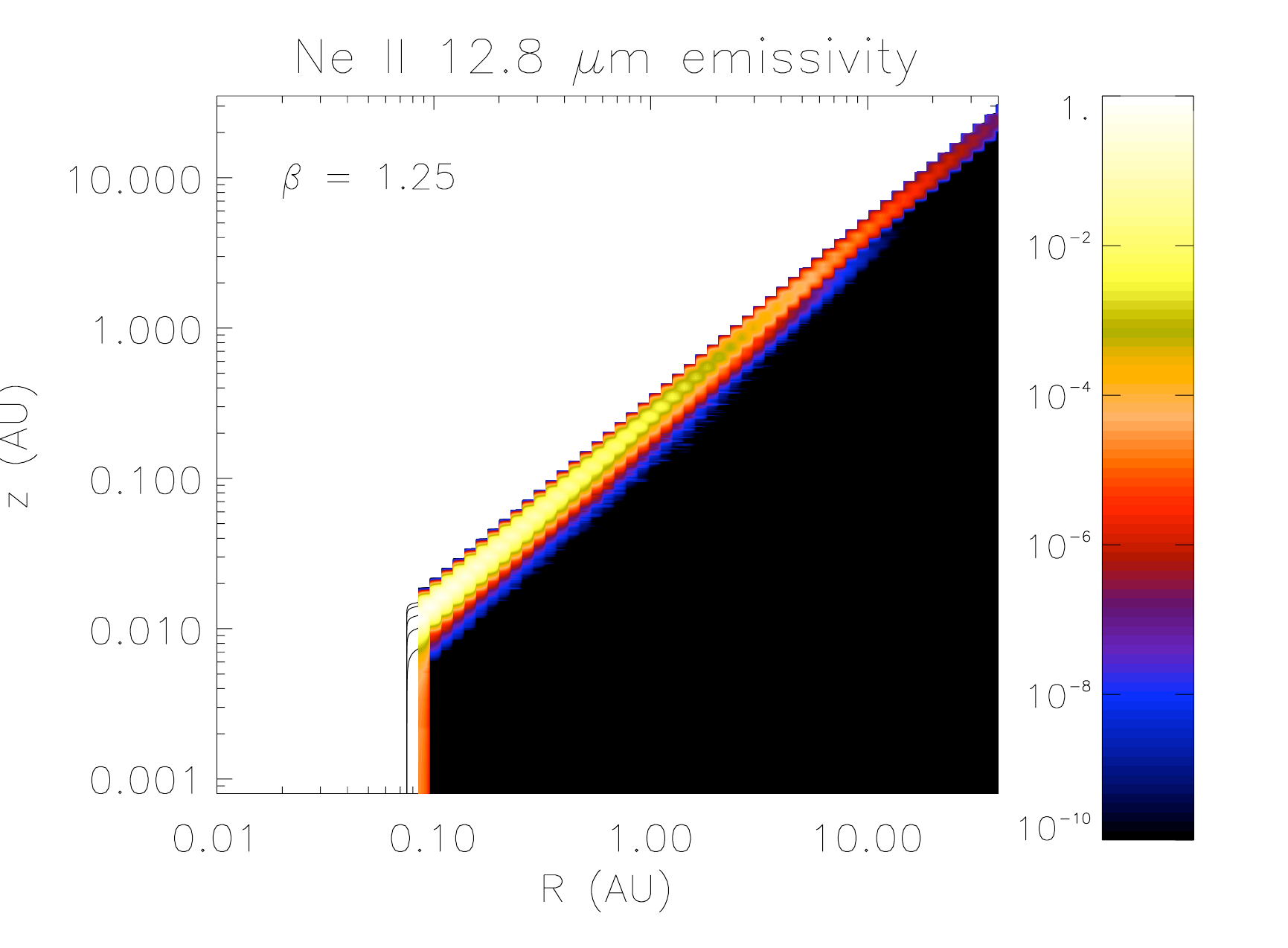}
\includegraphics[width=90mm,origin=c, trim = 25mm 125mm 15mm 25mm, clip]{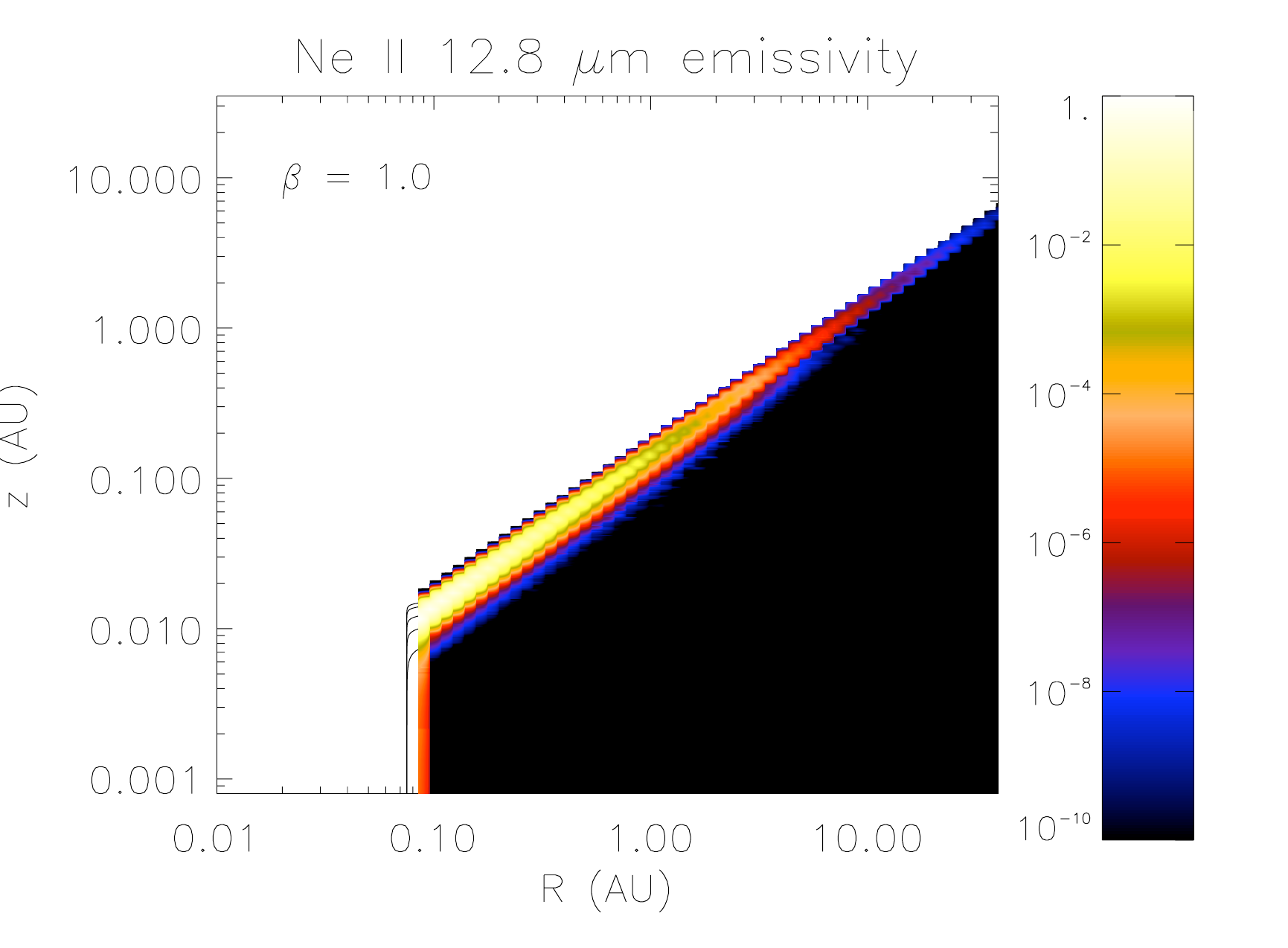}
\end{minipage}
\end{center}
\caption{Two-dimensional distribution of the [Ne~{\sc ii}] emissivity in arbitrary units. The {\it left} panel
refers to the disk model with $\beta$ = 1.25 (flared disk), instead the {\it right} panel show the disk model with
$\beta$=1.0 (flat disk).
}
\label{f:betaemission}
\end{figure*}

\begin{figure*}
\begin{center}
\centering
\begin{minipage}{.98\linewidth}
\hspace{-5mm}
\includegraphics[width=0.48\textwidth, trim = 25mm 125mm 15mm 25mm, clip]{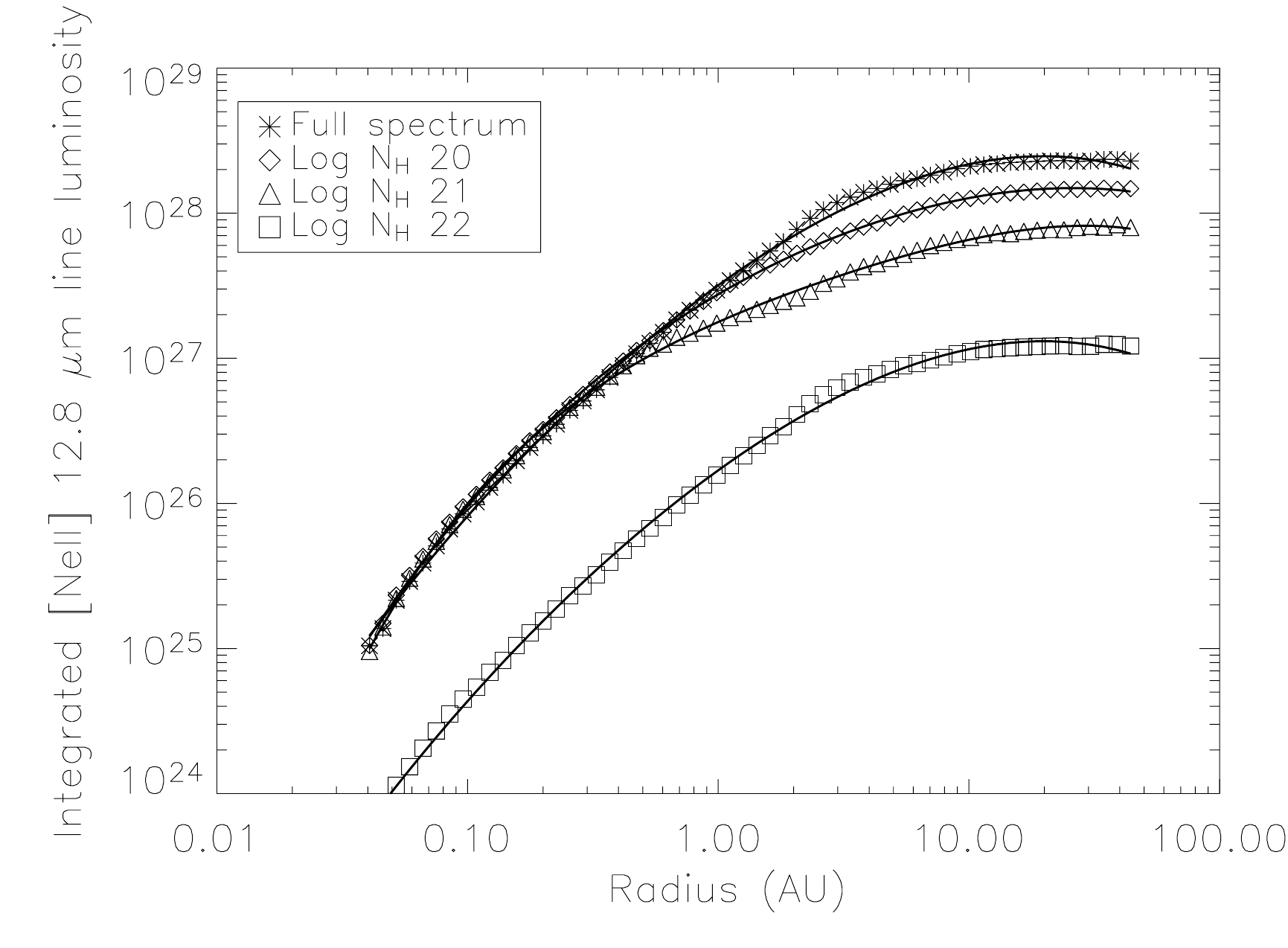}
\includegraphics[width=0.48\textwidth, trim = 25mm 125mm 15mm 25mm, clip]{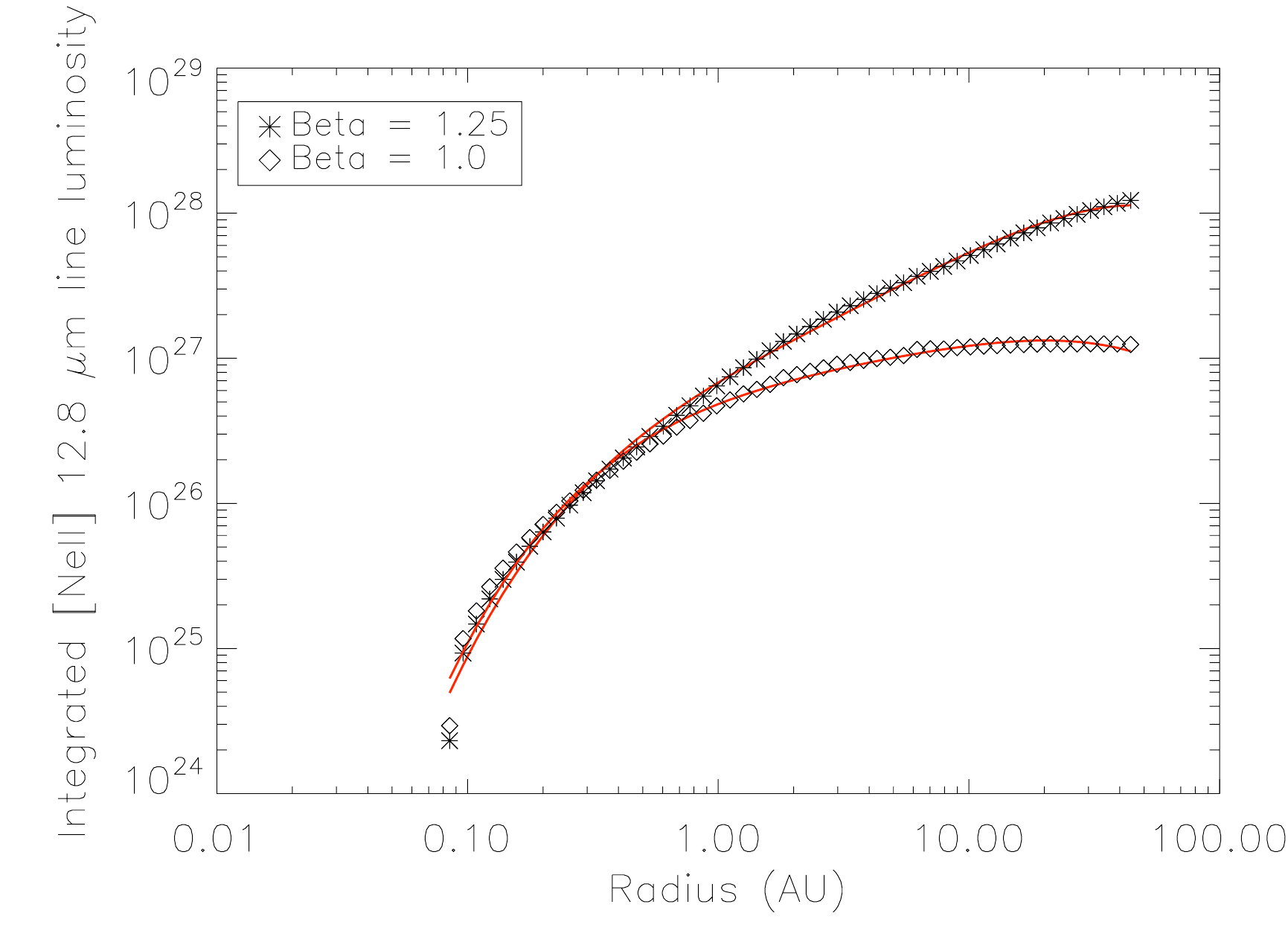}
\end{minipage}
\caption{[Ne~{\sc ii}] cumulative surface emission in erg s$^{-1}$(see text for details). {\it Left:} Hydrostatic equilibrium models irradiated by spectra screened by various neutral hydrogen columns. {\it Right:} Flat and flared disk models computed assuming Robitaille et al. (2006) density distribution.}
\label{Fig:IntegEmiss}
\end{center}
\end{figure*}

\begin{figure}
\centering
\includegraphics[width=0.48\textwidth, trim = 25mm 125mm 15mm 25mm, clip]{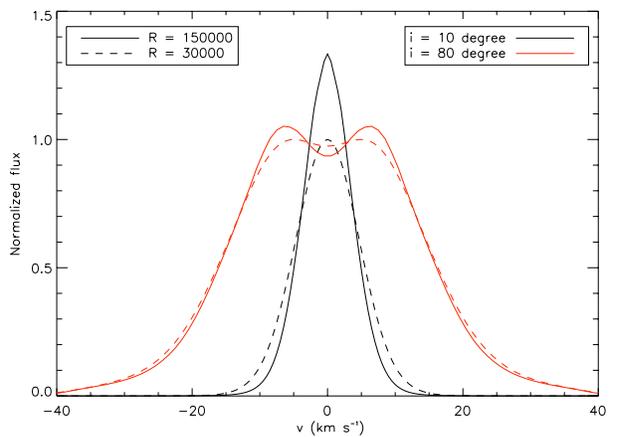}
\caption{[Ne~{\sc ii}] line profiles of a Keplerian disk in hydrostatic equilibrium irradiated by an unscreened EUV+X-ray source. The black lines are for low inclination (i = 10$^{\rm o}$) and the red lines for high inclination 
(i = 80$^{\rm o}$) disks. The solid lines have a resolution of R = 150000 and the dashed are the profiles degraded to a resolution of 30000. 
For clarity the lines are normalized to the peak value of the degraded case.
}
\label{Fig:lineFS0H0}
\end{figure}

\begin{figure*}
\centering
\begin{minipage}{.98\linewidth}
\hspace{-5mm}
\includegraphics[width=0.48\textwidth, trim = 25mm 125mm 15mm 25mm, clip]{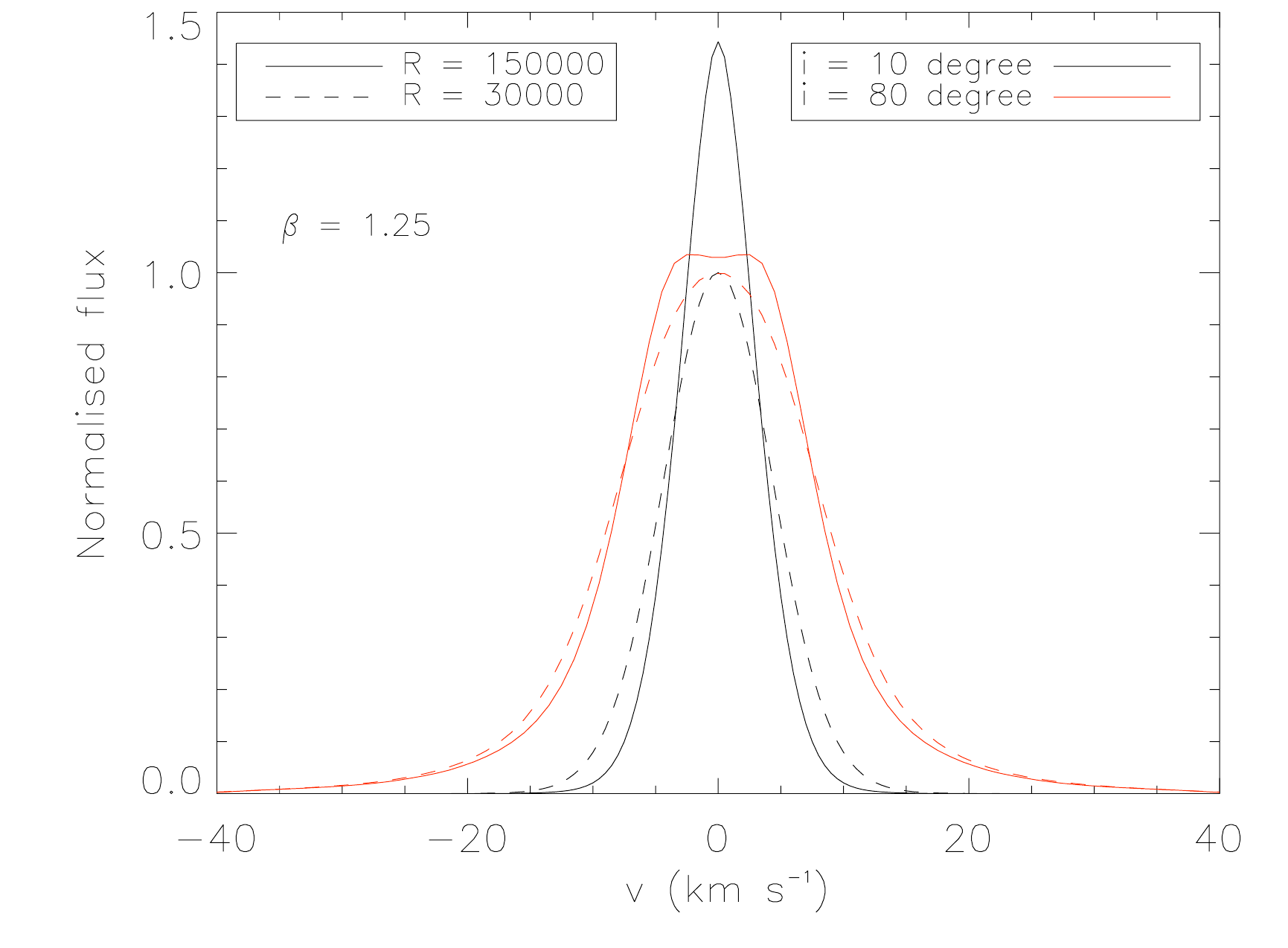}
\includegraphics[width=0.48\textwidth, trim = 25mm 125mm 15mm 25mm, clip]{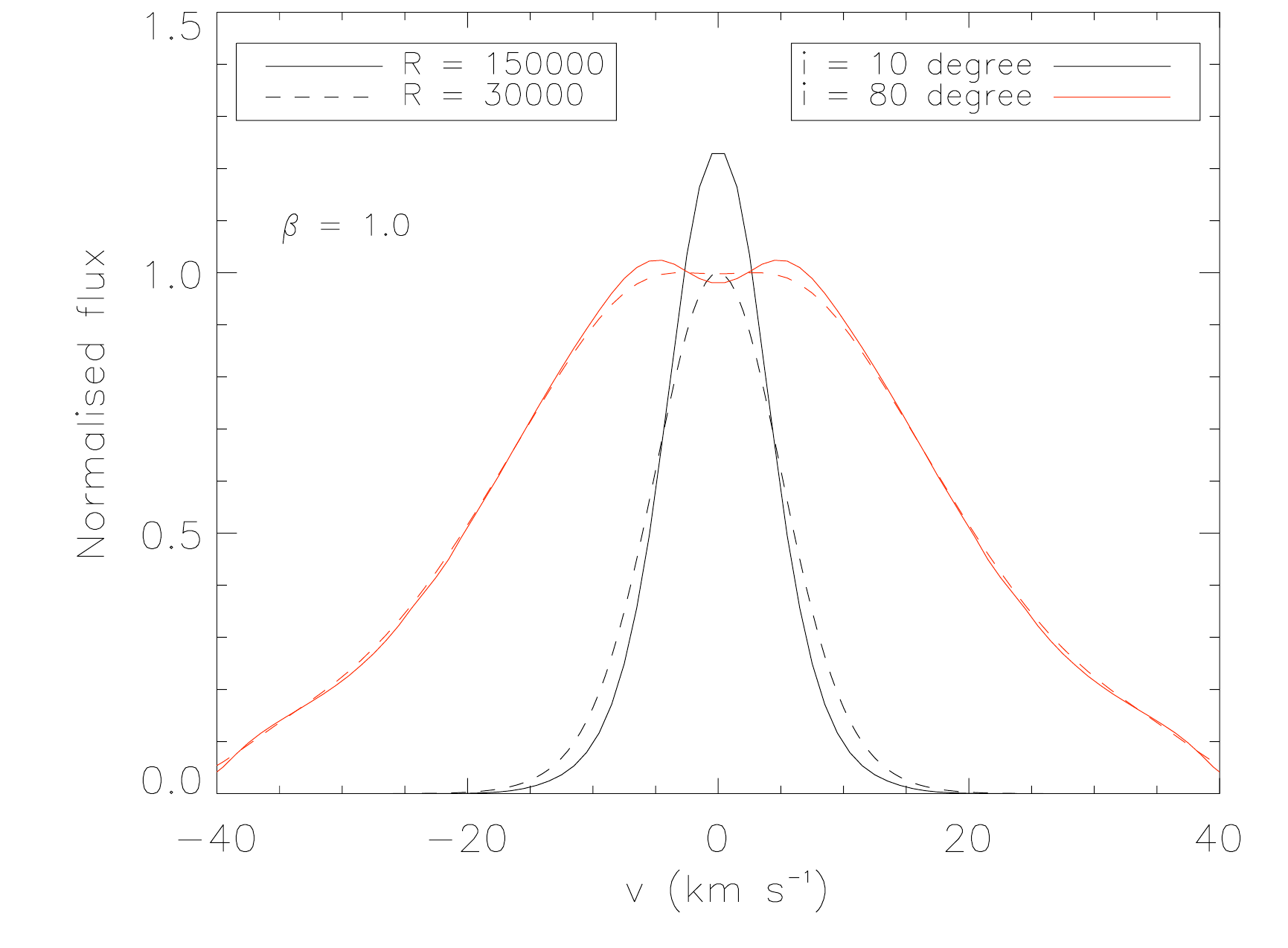}
\end{minipage}

\caption{As in Figure \ref{Fig:lineFS0H0}, but for the flared disk model ($\beta = 1.25$) ({\it left}) and flat disk model ($\beta = 1.00$) ({\it right}) irradiated by an unscreened EUV+X-ray spectrum.}
\label{Fig:lineBeta}
\end{figure*}

Disk flaring has also a dramatic effect on the size of the [Ne~{\sc ii}] emitting region. Figure~\ref{f:betaemission} shows two-dimensional plots of the [Ne~{\sc ii}]
emissivity for the flared and flat disk models, while the right panel of Figure~\ref{Fig:IntegEmiss} shows the surface-integrated emission for the line for the two disk
models. As summarised in Table~\ref{Table2}, the emission region in the flared disk model extends out to a radius of $\sim$30~AU, within which  90\% of the total [Ne~{\sc
ii}] luminosity is emitted. This radius is reduced to $\sim$6~AU for the flat disk model. 
Naturally, this has consequences on the line profiles for these two cases, which are shown in Figure~\ref{Fig:lineBeta} for two disk inclinations.
The FWHM for the [Ne~{\sc ii}] line produced by a flat disk seen almost edge on (inclination 80$^{\rm o}$) is roughly 40~km\, s$^{-1}$, while it is only 17\,km s$^{-1}$ for a flared disk where the
emission region extends to larger cylindrical radii. The difference is of course smaller if the disks are seen almost face on (inclination 10$^{\rm o}$), thus dominated by the contribution of the
thermal  broadening, where a FWHM of $\sim$10~km\, s$^{-1}$ is expected for a flat disk compared to a FWHM of $\sim$7~km\, s$^{-1}$ for a flared disk. These statistics are summarised in Table~\ref{Table3}. 

\begin{table}
\caption{[Ne~{\sc ii}] and [Ne~{\sc iii}] luminosities of disk irradiated by different high-energetic spectrum of L$_{X} (0.1-10 {\rm keV})$ = 2$\times$10$^{30}$ erg s$^{-1}$. 
The disk model adopted is a d'Alessio disk model, see also Ercolano et al. 2009 for details. }
\hbox{

\begin{tabular}{ ||l || c  c   c  ||}
\hline \hline
MODEL  &  L$_{\rm tot}$(13.6 eV-10 keV) & [Ne~{\sc ii}] &  [Ne~{\sc iii}]		  \\ 
& 10$^{30}$ erg s$^{-1}$ & \multicolumn{2}{c}{(10$^{28}$ erg s$^{-1})$} \\
\hline	
Unscreened & 4.0 &2.3& 4.9$\times10^{-1}$ \\
Log N$_h$ = 18& 2.3 & 1.9   & 4.3$\times10^{-1}$ \\
Log N$_h$ = 19& 2.04 & 1.7   & 3.2$\times10^{-1}$ \\
Log N$_h$ = 20& 2.0 & 1.5   & 2.2$\times10^{-1}$ \\
Log N$_h$ = 21& 2.0 & 9.1$\times10^{-1}$& 1.5$\times10^{-1}$ \\
Log N$_h$ = 22& 2.0  & 1.4$\times10^{-1}$& 2.5$\times10^{-2}$ \\
\hline
\hline

\label{Table1a}
\end{tabular}
}
\end{table}

\begin{table}
\caption{[Ne~{\sc ii}] line luminosities computed at a range of L$_{X}$. The corresponding values of L$_{tot}$ are also given.}
\begin{tabular}{ ||l || c   c c c  ||}

\hline \hline
MODEL  &    \multicolumn{3}{c}{[Ne~{\sc ii}]} 		  \\ 
&  \multicolumn{3}{c}{(10$^{28}$ erg s$^{-1})$} \\
\hline
L$_{X} [2\,10^{30}$erg\,s$^{-1}]$ & 0.1 & 1.0 & 10. \\
\hline

Unscreened (L$_{\rm tot} = 2\,{\rm L}_X$)     &   2.9$\times10^{-1}$ & 2.3    & 22.0     \\
Log(N$_H$)=21 (L$_{\rm tot} = {\rm L}_X$)  &   3.0$\times10^{-2}$ & 9.1$\times10^{-1}$ &  5.8  \\
\hline 
Flared (L$_{\rm tot} = 2\,{\rm L}_X$)& 1.0$\times10^{-1}$	 & 1.6    & 17.3 \\
Flat   (L$_{\rm tot} = 2\,{\rm L}_X$)    & 1.7$\times10^{-2}$	 & 1.3$\times10^{-1}$ & 7.0$\times10^{-1}$ \\
\hline \hline
\label{Table1b}
\end{tabular}
\end{table}

\begin{table}
\caption{Emissivity distribution statistics. See text for details.}
\begin{tabular}{ ||l | c   c ||}

\hline \hline

Model	  &  Half-L radius & 90\% L radius 		  \\ 
& (AU) & (AU) \\

\hline
Unscreened    & 2.9 & 10.1     \\
Log N$_h$= 20 & 3.3 & 13.0 		     \\
Log N$_h$= 21 & 3.5 & 12.8 		     \\
Log N$_h$= 22 & 2.9 & 10.1 		     \\

\hline
Flared $\beta$ = 1.25 &	12.9 & 34.3 \\
Flat $\beta$ = 1.0    &   1.5 & 6.1 \\
\hline 
\hline
\label{Table2}
\end{tabular}
\end{table}

\begin{table}
\caption{FWHM of the line profiles simulated for disks in Keplerian rotation viewed almost face on (10$^{\rm o}$ inclination) and almost edge on (80$^{\rm o}$ inclination) at two different spectral  resolutions.}
\begin{tabular}{ ||l | c  c  c c ||}
\hline \hline
	   &  \multicolumn{2}{c}{Inclination 10 degree}  & \multicolumn{2}{c}{Inclination 80 degree}  		  \\ 
	  &  \multicolumn{2}{c}{FWHM (km\,s$^{-1}$)}  & \multicolumn{2}{c}{FWHM (km\,s$^{-1}$)}  		  \\ 
\hline
R = & 150000 & 30000 & 150000 & 30000 \\

\hline
Unscreened & 8.2 & 11.1 & 31.3 & 32.6	             \\
Log N$_H$ = 21 & 6.2 & 10.0 & 25.8 & 28.1	     \\
\hline
Flared $\beta$ = 1.25 &	6.9 & 10.2 & 16.8 & 17.6 \\
Flat $\beta$ = 1.0    &   9.5 & 12.1 & 40.1 & 40.8 \\
\hline 
\hline
\label{Table3}
\end{tabular}
\end{table}

\section{Discussion}

As the sample of observed L([Ne~{\sc ii}]) luminosities from protoplanetary disks continues to grow, together with the sample of pre-main sequence stars with measured X-ray luminosities, a trend between
the two quantities becomes more and more evident (G\"udel et al. 2009). While this lends evidence to the [Ne~{\sc ii}] line being a product of X-ray irradiation of disks as predicted by Glassgold et
al. (2007), the large scatter associated with this relation makes the situation less clear-cut. For a handful of objects with known powerful outflows it is clear that this line cannot be used as
a diagnostics of the gaseous {\it disk} phase, as the emissivities may be dominated by gas in the outflows. However, even after removing objects with known outflows from our sample we are left with approximately {\it one order of magnitude variations in L([Ne~{\sc ii}]) at any given L$_X$}.

\begin{figure}
\centering
\includegraphics[width=9cm, trim = 25mm 125mm 15mm 25mm, clip]{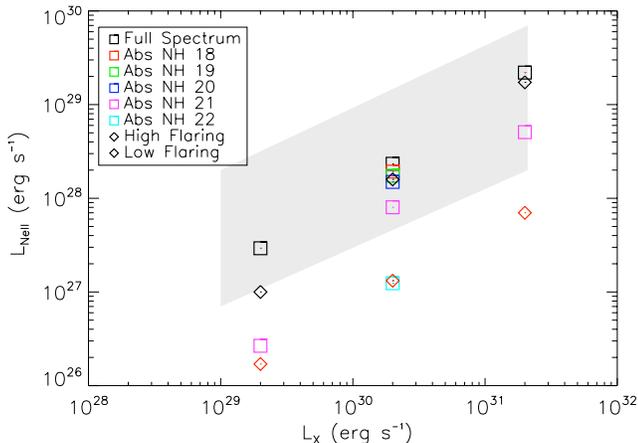}
\caption{Comparison between the prevision of our models, where a relationship between the L([Ne~{\sc ii}]) and L$_X$ is expected, and the observed system with undetected outflows (from
G\"udel 2008). }
\label{Fig:LNeIILX}
\end{figure}

In this paper we have investigated the origin of the large scatter observed in the [Ne~{\sc ii}] -- L$_X$ relation. We find that variations in the irradiating spectral shape and disk structure (flaring) are sufficient to explain the typically observed scatter of approximately one order of magnitude. 
Figure~\ref{Fig:LNeIILX} shows the results from our models (coloured symbols) overplotted on a shaded region representing roughly the observations presented by G\"udel et al. (2009), excluding the strong outflow objects (blue points in their plot).  The definition of such region is affected by the limited sample statistic that does not allow a sharper outline, but it shows where the bulk of data falls. 
Moreover, we note here that we use luminosities {\it after} the screen while G\"udel et al. use intrinsic luminosities derived from the observations. We have performed absorption calculations for a subset of the observations from which we found the following: absorbing the intrinsic X-ray spectral models by $N_{\rm H} = 3\times 10^{20}~{\rm cm}^{-2} - 4.6\times 10^{22}~{\rm cm}^{-2}$ (a range occupied by the observations) reduces the 0.1-10~keV luminosity by factors of approximately 1.4--5.5 (also depending on the intrinsic thermal model from the individual fits). On average, thus, the {\it unabsorbed} $L_{\rm X}$ from the observations should be shifted  by a factor of 3 (0.5~dex) to the left to obtain average absorbed values, with individual shifts ranging from 0.15~dex to 0.75~dex. These corrections do not change the plot  qualitatively. We, however, refrain from using the individual {\it absorbed} (post-screen) luminosities derived from the observations here because it is not clear a priori 
that the absorbed luminosity seen from Earth is close to the absorbed luminosity seen by the disk. The latter luminosity is principally unknown. 
We also mention that G\"udel et al. use the 0.3-10~keV range for their luminosities, while we use 0.1-10~keV. For the post-screen luminosities for stars with
$N_{\rm H} > 10^{20}$~cm$^{-2}$, however, the difference is minor.

Our model predictions seem to agree well with the observations in terms of reproducing the scatter, and at intermediate and high X-ray luminosities the absolute values of
L([Ne~{\sc ii}]) also roughly agree with the observations. Models with low X-ray luminosities, however, fall short of the observed values, and while detection limits can partially affect the observations, the failure of our models to produce L([Ne~{\sc ii}]) $>$ 10$^{28}erg/sec$ for X-ray luminosities of 2$\times$10$^{29}erg\,s^{-1}$ must have another origin. 
Nevertheless there is an overlap between our prediction for the unabsorbed source and/or the flared disk and the shaded region.
Keep in mind, however, that the correction for absorption mentioned above aggravates the discrepancy between the model predictions and the observations (the latter being moved somewhat to the left in Fig. 8), but do not have effects on the conclusions on the scatter of [Ne~{\sc ii}] luminosities. 

It would be of little use here to speculate what other observational effects may be coming into play, however one thing that is worth noticing is that the measurements 
plotted in the Figure all come from the {\it Spitzer} Space Telescope, while recent observations from the ground at higher resolution suggest often 
lower luminosities than the {\it Spitzer} data \citep{Herczeg2007,Najita2009}. This has been interpreted as possible pollution from molecular emission in the {\it Spitzer} 
band, while more recent works (Najita et al., 2009) seem to lean toward extended emission, from undetected outflows, which would contribute to the flux in the {\it Spitzer} 
aperture, but would be excluded by the narrow slits used for measurements from the ground. 

Our results thus support a disk origin of the observed [Ne~{\sc ii}] emission as a consequence of X-ray irradiation, at least for systems with moderate [Ne~{\sc ii}]  luminosities and 
absence of jets. On the other hand, the additional parameters (X-ray spectral hardness, disk flaring) now found to influence the [Ne~{\sc ii}] luminosity add complexity to the interpretation, 
and the usefulness of the [Ne~{\sc ii}] line as a disk diagnostic depends on our ability to confine these other parameters as well. Clearly, [Ne~{\sc ii}]  {\it disk} diagnostics needs to 
avoid stars with known jets and outflows. Under the assumption of full dust and gas mixing, some information about disk flaring can be obtained by modeling the infrared spectral
energy distribution \citep{Chiang2001,Pascucci2003}, providing some important constraints on the range of modeled [Ne~{\sc ii}]  luminosities.
Constraining the shape of the ionizing spectrum is perhaps the most difficult task. The intrinsic spectrum can usually be modeled sufficiently well based on intermediate- or high-resolution 
X-ray spectroscopy, but the absorbing gas column density toward the disk may be different from that toward the observer.
However, gas absorption matters the least in systems with weak or absent accretion, the possible presence of inner holes, and therefore they likely absence of strong winds emanating 
from the innermost disk regions. The [Ne~{\sc ii}]  {\it disk } diagnostic may thus be optimized for the study of transitional disks, most of which are also not known to drive outflows or
jets.  Several of these have been well studied, including new ground-based observations of spectrally resolved  [Ne~{\sc ii}] lines (Najita et al. 2009). In this context, the [Ne~{\sc ii}] line may develop its full diagnostic power to study disk ionization and heating by ionizing radiation from the central star, processes that are pivotal for disk dispersal through photoevaporation \citep{Alexander2006, Ercolano2008a, Ercolano2009, Owen2009}in the first place.

High resolution and high signal-to-noise line profile can help assessing the origin of the line, with profiles centred at the stellar radial velocity being synonymous with a disk origin. This is true for all inclinations except those very close to edge on, where outflows would also produce profiles centered on the stellar radial velocity. Blue shifted profiles are expected to be observed for system with outflows observed at non edge on inclinations (e.g. Alexander, 2008), and have been recently oberved by Pascucci et al. (2009) in the spectra of TW~Hya, Sz~73, T~Cha and CS~Cha. 
Other examples of high resolution spectra where the line has been detected include TW Hya (Herczeg et al. 2007; $\lambda/\Delta\lambda \sim$ 30000), GM Aur and
AA Tau (Najita et al., 2009; $\lambda/\Delta\lambda \sim$ 80000). Here the observed lines are consistent with being centred at the stellar radial velocity, suggesting a circumstellar disk origin, although the poor signal-to-noise of some of these detections makes it hard to say with certainty. The lines are broad with a FWHM of respectively  $\sim$ 21 (TW Hya), 70 (AA Tau) and 14 (GM Aur) km s$^{-1}$. The [Ne~{\sc ii}] FWHM of GM Aur can be produced by a disk of normal flaring power, however the large FWHM of AA Tau requires the emission region to be dominated by very small radii and therefore implies a small degree of flaring. 
We finally conclude that high resolution observations of the [Ne~{\sc ii}] line in YSOs, able to resolve its profile, are needed if this line is to be used to extract useful information on disk structure (e.g. flaring) and evolutionary stage by (e.g.) the detection of outflows and photoevaporative winds by comparison with line profile models like those of Alexander (2008) or those presented in this work.

\section*{Acknowledgments}
We thank Juan Alcal\'a, Elvira Covino, Al Glassgold and Joan Najita for the useful comments. We further thank Thomas Robitaille for the information on the density structure of his disk model. 
This research was supported financially  from INAF (PRIN 2007: {\em From active accretion to debris discs}).  


\label{lastpage}

\end{document}